\documentclass{article}
\usepackage{amsfonts}
\usepackage{amsmath}
\usepackage{graphicx}

\begin{document}

\title{Chaotic dynamics in a simple bouncing ball model}
\author{Andrzej Okninski$^{1}$, Bogus\l aw Radziszewski$^{2}$ \\
Physics Division$^{1}$, Department of Mechatronics and Mechanical Engineering%
$^{2}$, \\
Politechnika Swietokrzyska, Al. 1000-lecia PP 7, 25-314 Kielce, Poland}
\date{}
\maketitle

\begin{abstract}
We study dynamics of a ball moving in gravitational field and colliding with
a moving table. The motion of the limiter is assumed as periodic with
piecewise constant velocity -- it is assumed that the table moves up with a
constant velocity and then moves down with another constant velocity. The
Poincar\'{e} map, describing evolution from an impact to the next impact, is
derived and scenarios of transition to chaotic dynamics are investigated analytically.
\end{abstract}

\section{Introduction}

Vibro-impacting systems belong to a very interesting and important class of
nonsmooth and nonlinear dynamical systems \cite%
{diBernardo2008,Luo2006,Awrejcewicz2003,Filippov1988}\ with important
technological applications \cite%
{Stronge2000,Mehta1994,Knudsen1992,Wiercigroch2008}. Dynamics of such
systems can be extremely complicated due to velocity discontinuity arising
upon impacts. A very characteristic feature of such systems is the presence
of nonstandard bifurcations such as border-collisions and grazing impacts
which often lead to complex chaotic motions.

The Poincar\'{e} map, describing evolution from an impact to the next
impact, is a natural tool to study vibro-impacting systems. The main
difficulty with investigating impacting systems is in finding instant of the
next impact what typically involves solving a nonlinear equation. However,
the problem can be simplified in the case of a bouncing ball dynamics
assuming a special motion of the limiter. In the present paper we
investigate motion of a material point in a gravitational field colliding
with a limiter moving with piecewise constant velocity. This class of models
has been extensively studied, see \cite{Luo2009}\ and references therein. As
a motivation that inspired this work, we mention study of physics and
transport of granular matter \cite{Mehta1994}. A similar model has been also
used to describe the motion of railway bogies \cite{Knudsen1992}. Therefore
it can be expected that some of the present results may cast light on the
dynamics in such systems. On the other hand, simple motion of the limiter
makes analytical explorations possible, cf. our preliminary report \cite%
{Okninski2009}.

The paper is organized as follows. In Section 2 a one dimensional dynamics
of a ball moving in a gravitational field and colliding with a table is
considered and Poincar\'{e} map is described for piecewise linear motion of
the table. In Section 3 transition to chaotic dynamics from periodic motion
is described. The nature of mixing leading to chaotic dynamics is described
in Section 4. In Sections 5 and 6 homoclinic structures responsible for
mixing are determined and computed. Finally, we discuss our results in the
last Section.

\section{Bouncing ball: a simple motion of the table}

We consider a motion of a small ball moving vertically in a gravitational
field and colliding with a moving table, representing unilateral
constraints. The ball is treated as a material point while the limiter's
mass is assumed so large that its motion is not affected at impacts. A
motion of the ball between impacts is described by the Newton's law of
motion:%
\begin{equation}
m\ddot{x}=-mg,  \label{point motion}
\end{equation}%
where $\dot{x}=dx/dt$ and motion of the limiter is:%
\begin{equation}
y=y\left( t\right) ,  \label{limiter motion}
\end{equation}%
with a known function $y$. We shall also assume that $y$ is a continuous
function of time. Impacts are modeled as follows: 
\begin{eqnarray}
x\left( \tau _{i}\right) &=&y\left( \tau _{i}\right) ,  \label{position} \\
\dot{x}\left( \tau _{i}^{+}\right) -\dot{y}\left( \tau _{i}\right)
&=&-R\left( \dot{x}\left( \tau _{i}^{-}\right) -\dot{y}\left( \tau
_{i}\right) \right) ,  \label{velocity}
\end{eqnarray}%
where duration of an impact is neglected with respect to time of motion
between impacts. In Eqs. (\ref{position}), (\ref{velocity}) $\tau _{i}$
stands for time of the $i$-th impact while $\dot{x}_{i}^{-}$, $\dot{x}%
_{i}^{+}$are left-sided and right-sided limits of $\dot{x}_{i}\left(
t\right) $ for $t\rightarrow \tau _{i}$, respectively, and $R$ is the
coefficient of restitution, $0\leq R<1$ \cite{Stronge2000}.

Solving Eq. (\ref{point motion}) and applying impact conditions (\ref%
{position}), (\ref{velocity}) we derive the Poincar\'{e} map \cite{AOBR2007}%
: 
\begin{equation}
\begin{array}{l}
\gamma Y\left( T_{i+1}\right) =\gamma Y\left( T_{i}\right) -\Delta
_{i+1}^{2}+\Delta _{i+1}V_{i}, \\ 
V_{i+1}=-RV_{i}+2R\Delta _{i+1}+\gamma \left( 1+R\right) \dot{Y}\left(
T_{i+1}\right) ,%
\end{array}
\label{YV}
\end{equation}%
where $\Delta _{i+1}\equiv T_{i+1}-T_{i}$. The limiter's motion has been
typically assumed in form $Y(T)=\sin (T)$, cf. \cite{AOBR2009} and
references therein. This choice leads to serious difficulties in solving the
first of Eqs.(\ref{YV}) for $T_{i+1}$, thus making analytical investigations
of dynamics hardly possible. Accordingly, we have decided to simplify the
limiter's periodic motion to make (\ref{YV}) solvable. Let us thus assume
that the table moves up with a finite constant velocity $\gamma \dot{Y}_{1}$
and then goes down with a finite constant velocity $\gamma \dot{Y}_{2}$ \cite%
{AOBR2009}. Therefore, displacement of the table is the following periodic
function of time: 
\begin{equation}
Y\left( T\right) =\left\{ 
\begin{array}{l}
\frac{1}{h}\left( T-\left\lfloor T\right\rfloor \right) ,\quad \left(
T-\left\lfloor T\right\rfloor <h\right) \\ 
\frac{-1}{1-h}\left( T-\left\lfloor T\right\rfloor \right) +\frac{1}{1-h}%
,\quad \left( h<T-\left\lfloor T\right\rfloor \right)%
\end{array}%
\right.  \label{displacement}
\end{equation}%
with $\dot{Y}_{1}=1/h$, $\dot{Y}_{2}=-1/\left( 1-h\right) $, $0<h<1$, where $%
\left\lfloor x\right\rfloor $ is the floor function -- the largest integer
less than or equal to $x$. Our model consists thus of equations (\ref{YV}), (%
\ref{displacement}) with control parameters $R$, $\gamma $, $h$. Since the
period of motion of the limiter is equal to one, the map (\ref{YV}) is
invariant under the translation $T_{i}\rightarrow T_{i}+1$. Accordingly, all
impact times $T_{i}$ can be reduced to the unit interval $\left[ 0,\ 1\right]
$.

\section{From periodic dynamics to chaotic motion}

In our recent article periodic solutions of Eqs. (\ref{YV}), (\ref%
{displacement}) have been investigated \cite{AOBR2009}. Dynamics of Eqs. ( %
\ref{YV}), (\ref{displacement}) becomes complicated when some impacts occur
in time interval $\left( 0,h\right) $, and some in $\left( h,1\right) $.

In the case of two impacts per period a $2$ - cycle, $T_{1}\in \left(
0,\,h\right) $, $T_{2}\in \left( h,\,1\right) $, is stable. Next, for
increasing values of $\gamma $, period doubling takes place and $2^{2}$ -
cycle with impacts $0<T_{\ast 1},T_{\ast 3}<h$, $h<T_{\ast 2},T_{\ast 4}<1$
is formed. Then, upon further increase of $\gamma $, the period doubling
scenario ends abruptly when $T_{\ast 2}=1$. For $R=0.85,\,h=0.2623$ this
happens for $\gamma _{cr}^{\left( 1\right) }=0.09058194712$. This critical
transition is investigated in Sections 4, 5. Equations for dynamics after
the first period doubling have the following form:

\begin{equation}
\left\{ 
\begin{array}{l}
\frac{\gamma \left( -T_{2}+1\right) }{1-h}=\frac{\gamma }{h}T_{1}-\left(
T_{2}-T_{1}\right) ^{2}+\left( T_{2}-T_{1}\right) V_{1} \\ 
V_{2}=-RV_{1}+2R\left( T_{2}-T_{1}\right) -\frac{\gamma \left( 1+R\right) }{%
1-h} \\ 
\frac{\gamma \left( T_{3}-1\right) }{h}=\frac{\gamma \left( -T_{2}+1\right) 
}{1-h}-\left( T_{3}-T_{2}\right) ^{2}+\left( T_{3}-T_{2}\right) V_{2} \\ 
V_{3}=-RV_{2}+2R\left( T_{3}-T_{2}\right) +\frac{\gamma \left( 1+R\right) }{h%
} \\ 
\frac{\gamma \left( -T_{4}+1\right) }{1-h}=\frac{\gamma T_{3}}{h}-\left(
T_{4}-T_{3}\right) ^{2}+\left( T_{4}-T_{3}\right) V_{3} \\ 
V_{4}=-RV_{3}+2R\left( T_{4}-T_{3}\right) +\frac{\gamma \left( 1+R\right) }{%
1-h} \\ 
\frac{\gamma T_{5}}{h}=-\frac{\gamma T_{5}}{1-h}-\left( T_{5}-T_{4}\right)
^{2}+\left( T_{5}-T_{4}\right) V_{4} \\ 
V_{5}=-RV_{4}+2R\left( T_{5}-T_{4}\right) +\frac{\gamma \left( 1+R\right) }{h%
} \\ 
T_{5}=T_{1}+1 \\ 
V_{5}=V_{1}%
\end{array}%
\right.  \label{equations}
\end{equation}

\begin{figure}[!h]
\begin{equation*}
\includegraphics[width=8 cm]{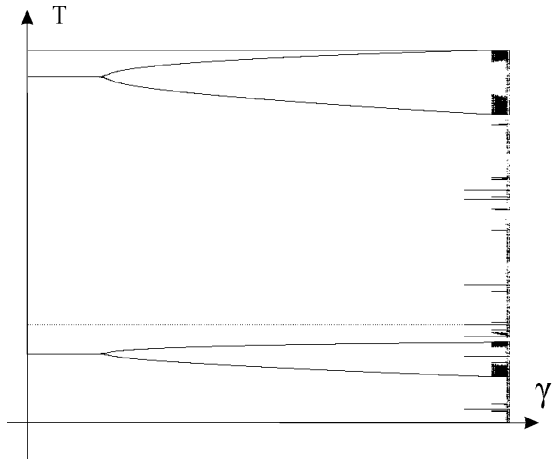}
\end{equation*}%
\caption{Bifurcation diagram: $R=0.85$, $h=0.2623$, $\protect\gamma \in %
\left[ 0.09,0.090605\right] $.}
\label{F1}
\end{figure}

In Fig. 1 transition to chaos is shown. The initial dynamical state with two
impacts per period bifurcates at $\gamma =\gamma _{pd}^{\left( 2\right)
}=0.090099$ (this value can be computed analytically, see \cite{AOBR2009}).
For $\gamma \rightarrow \gamma _{cr}^{\left( 2\right) }$ time of the second
impact tends to $1$ and this mode of dynamics is impossible for $\gamma
>\gamma _{cr}^{\left( 2\right) }$. It turns out that for $\gamma >\gamma
_{cr}^{\left( 2\right) }$ there are two attractors: a noisy, probably
chaotic, attractor coexisting with a $7$--cycle which appears just before
the transition. At $\gamma =0.09060$ the noisy attractor disappears and is
substituted by a more irregular attractor, see Figs. 2,3.

Full circles indicate positions of small clouds of points. We have studied
these potentially chaotic attractors in detail. First of all, we have
checked numerically that the attractors are non--periodic. Indeed,
computations show that after $10^{8}$ iterations the points generated by the
map (\ref{YV}), (\ref{displacement}) stay on the corresponding attractor and
do not repeat. We have also computed Lyapunov exponents for both attractors.
In the case of the attractor shown in Fig. 2 the Lyapunov exponent is $%
\lambda _{1}=0.5$ while for the attractor in Fig. 3 $\lambda _{2}=0.8$. It
follows that in both cases dynamics is chaotic and is more mixing in the
second case. The mechanism of mixing is explained in the next Section.

\begin{figure}[!h]
\begin{equation*}
\includegraphics[width=7 cm]{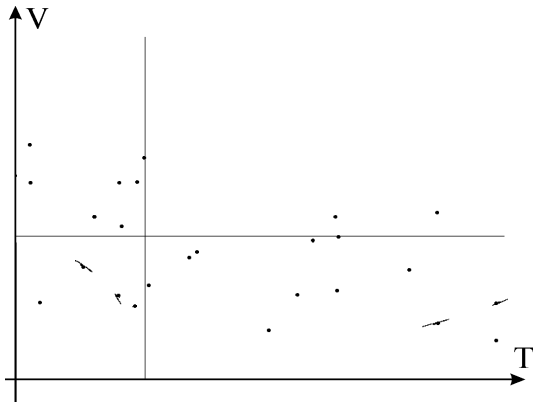}
\end{equation*}%
\caption{Chaotic attractor, $R=0.85,h=0.2623,\protect\gamma =0.09059$.}
\label{F2}
\end{figure}

\begin{figure}[h]
\begin{equation*}
\includegraphics[width=7 cm]{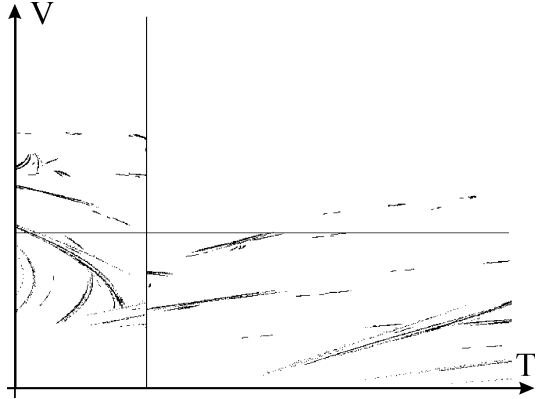}
\end{equation*}%
\caption{Chaotic attractor, $R=0.85,h=0.2623,\protect\gamma =0.09060$.}
\end{figure}

\newpage 

\section{Mechanism of mixing}

Mixing can arise due to corner events \cite{diBernardo2008} when impacts
occur at points where motion of the limiter loses smoothness at time
instances $T_{cr}^{\left( 1\right) }=h$, $T_{cr}^{\left( 2\right) }=1$. Let
us investigate the second possibility more closely. In Fig. 4 the stable $%
2^{2}$ - cycle with four impacts per two periods: $0<T_{\ast 1},T_{\ast 3}<h$%
, $h<T_{\ast 2},T_{\ast 4}<1$ and unstable $2$ - cycle are shown
schematically.

\begin{figure}[h]
\begin{equation*}
\includegraphics[width=7 cm]{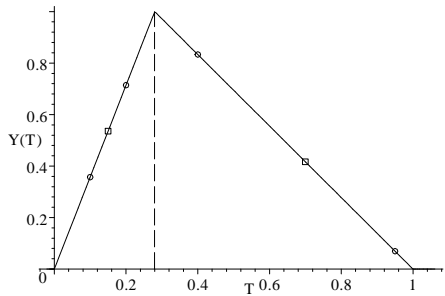}
\end{equation*}%
\caption{Stable $2^{2}$ - cycle (circles) and the unstable $2$ - cycle
(boxes).}
\label{F4}
\end{figure}

For increasing value of the control parameter $\gamma $ we have $T_{\ast
2}\rightarrow T_{cr}^{\left( 2\right) }=1$, see Fig. 1. The map (\ref{YV}), (%
\ref{displacement}) is invariant under translation $T_{i}\rightarrow T_{i+1}$
and thus the phase space is topologically equivalent to the cylinder and
hence we have to glue the end points of the time interval $\left[ 0,1\right] 
$ obtaining thus a circle. Therefore, a small neighborhood of $%
T_{cr}^{\left( 2\right) }=1$ is a union of two sets, $O_{cr}^{\left(
2\right) }=\left\{ T:\quad 1-\epsilon _{1}<T\leq 1\right\} \cup \left\{
T:\quad 0\leq T<\epsilon _{2}\right\} $, where $\epsilon _{1},\ \epsilon
_{2} $ are small and positive, see Fig. 5.

\begin{figure}[h]
\begin{equation*}
\includegraphics[width=7 cm]{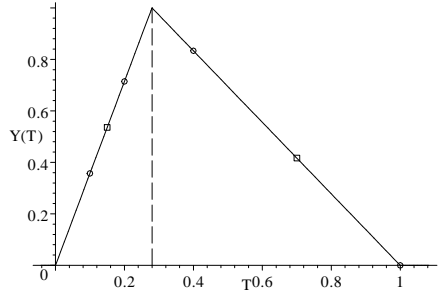}
\end{equation*}%
\caption{Unstable $2^{2}$ - cycle (circles) and the unstable $2$ - cycle
(boxes).}
\label{F5}
\end{figure}

Now, let $T_{i}\in O_{cr}^{\left( 2\right) }$. It follows from Eqs. (\ref{YV}%
), (\ref{displacement}) that time of the next impact, $T_{i+1}$, as well as
the corresponding post--impact velocity, $V_{i+1}$, depend discontinuously
on $T_{i}$. In other words, we get different solutions, $T_{i+1},\,V_{i+1}$,
depending on whether $1-\epsilon _{1}<T_{i}<1$ or $0<T_{i}<\epsilon _{2}$.
It follows that mixing will necessarily be present if a trajectory
recurrently visits the interval $O_{cr}^{\left( 2\right) }$. We shall study
this possibility in the next Section.

\section{Origin of the homoclinic structure}

Let us consider critical case: $\gamma =\gamma _{cr}^{\left( 1\right) }$, $%
T_{\ast 2}=T_{cr}^{\left( 2\right) }=1$. This is described by Eq. (\ref%
{equations}) with four impacts per two periods such that $0<T_{\ast
1},T_{\ast 3}<h$, $T_{\ast 2}=1,T_{\ast 4}<1$:%
\begin{equation}
\left\{ 
\begin{array}{l}
0=\frac{\gamma }{h}T_{1}-\left( 1-T_{1}\right) ^{2}+\left( 1-T_{1}\right)
V_{1} \\ 
V_{2}=-RV_{1}+2R\left( 1-T_{1}\right) -\frac{\gamma \left( 1+R\right) }{1-h}
\\ 
\frac{\gamma }{h}=-\left( T_{3}-1\right) +V_{2} \\ 
V_{3}=-RV_{2}+2R\left( T_{3}-1\right) +\frac{\gamma \left( 1+R\right) }{h}
\\ 
\frac{\gamma \left( -T_{4}+1\right) }{1-h}=\frac{\gamma T_{3}}{h}-\left(
T_{4}-T_{3}\right) ^{2}+\left( T_{4}-T_{3}\right) V_{3} \\ 
V_{4}=-RV_{3}+2R\left( T_{4}-T_{3}\right) +\frac{\gamma \left( 1+R\right) }{%
1-h} \\ 
\frac{\gamma T_{1}}{h}=-\frac{\gamma T_{1}}{1-h}-\left( T_{1}+1-T_{4}\right)
^{2}+\left( T_{1}+1-T_{4}\right) V_{4} \\ 
V_{1}=-RV_{4}+2R\left( T_{1}+1-T_{4}\right) +\frac{\gamma \left( 1+R\right) 
}{h}%
\end{array}%
\right.  \label{hom1}
\end{equation}%
where we have substituted $T_{2}=1$ and the periodicity conditions: $%
T_{5}=T_{1}+1$, $V_{5}=V_{1}$. Solutions of Eq. (\ref{hom1}) depend on roots
of the following algebraic equation:%
\begin{equation}
A_{6}X^{6}+A_{5}X^{5}+A_{4}X^{4}+A_{3}X^{3}+A_{2}X^{2}+A_{1}X+A_{0}=0,
\label{poly}
\end{equation}%
where $X\equiv T_{1}$ with coefficients $A_{i}$ depending on $R$, $h$ in a
complicated way. The coefficients are listed below:%
\begin{equation*}
A_{6}=\left( R+1\right) ^{2}\left( R^{2}+1\right) ^{5},
\end{equation*}%
$A_{5}=a_{5}+b_{5}h$:%
\begin{eqnarray*}
a_{5} &=&2\left( R+1\right) \left( 
\begin{array}{c}
R^{11}-11R^{10}-29R^{9}-17R^{8}+16R^{7}-36R^{6} \\ 
-48R^{5}-36R^{4}+23R^{3}-9R^{2}-11R-3%
\end{array}%
\right) , \\
b_{5} &=&-4R\left( R^{3}-R-2\right) \left( R+1\right) ^{2}\left(
R^{2}+1\right) ^{3},
\end{eqnarray*}%
$A_{4}=a_{4}+b_{4}h+c_{4}h^{2}$:%
\begin{eqnarray*}
a_{4} &=&23+74R-6R^{2}-158R^{3}+121R^{4}+348R^{5}+316R^{6} \\
&&+28R^{7}+73R^{8}+250R^{9}+170R^{10}+34R^{11}+7R^{12}, \\
b_{4} &=&4\left( R+1\right) \left( 
\begin{array}{c}
17R^{10}+11R^{9}-20R^{8}-49R^{7}+5R^{6} \\ 
-19R^{5}-41R^{4}-55R^{3}+10R^{2}-3%
\end{array}%
\right) , \\
c_{4} &=&4\left( R^{4}+1\right) \left( R+1\right) ^{2}\left( R^{2}+1\right)
^{3},
\end{eqnarray*}%
$A_{3}=a_{3}+b_{3}h+c_{3}h^{2}$:%
\begin{eqnarray*}
a_{3} &=&4\left( R+1\right) \left( 
\begin{array}{c}
R^{11}-11R^{10}-41R^{9}-55R^{8}+28R^{7}-32R^{6} \\ 
-48R^{5}-32R^{4}+3R^{3}+43R^{2}-7R-9%
\end{array}%
\right) , \\
b_{3} &=&16-72R-160R^{2}+352R^{3}+520R^{4}+336R^{5}+32R^{6} \\
&&+416R^{7}+512R^{8}+184R^{9}-128R^{10}-64R^{11}-24R^{12}, \\
c_{3} &=&-8R\left( R+1\right) \left( R^{4}+1\right) \left(
R^{6}+9R^{5}+2R^{4}-R^{2}+15R-2\right) ,
\end{eqnarray*}%
$A_{2}=a_{2}+b_{2}h+c_{2}h^{2}+d_{2}h^{3}$:%
\begin{eqnarray*}
a_{2} &=&31-2R-118R^{2}-178R^{3}+9R^{4}+188R^{5}+252R^{6} \\
&&-180R^{7}+113R^{8}+358R^{9}+154R^{10}+6R^{11}+7R^{12}, \\
b_{2} &=&-8+192R+120R^{2}-168R^{3}-416R^{4}-176R^{5}-192R^{7} \\
&&-632R^{8}-208R^{9}+72R^{10}+40R^{11}-32R^{12}, \\
c_{2} &=&8\left( R^{4}+1\right) \left( 
\begin{array}{c}
6R^{8}+10R^{7}+11R^{6}-8R^{5} \\ 
+3R^{4}+14R^{3}+21R^{2}-8R-1%
\end{array}%
\right) , \\
d_{2} &=&16R\left( R-1\right) \left( R+1\right) \left( R^{4}+1\right) ^{2},
\end{eqnarray*}%
$A_{1}=a_{1}+b_{1}h+c_{1}h^{2}+d_{1}h^{3}$:%
\begin{eqnarray*}
a_{1} &=&-2\left( R+1\right) ^{2}\left( 
\begin{array}{c}
3R^{10}+4R^{9}-3R^{8}+92R^{7}-132R^{6} \\ 
+80R^{5}-16R^{4}+28R^{3}-47R^{2}+4R+3%
\end{array}%
\right) , \\
b_{1} &=&-32-64R-180R^{2}+56R^{3}+204R^{4}+112R^{5}-280R^{6} \\
&&+128R^{7}+392R^{8}+80R^{9}-116R^{10}+72R^{11}+12R^{12}, \\
c_{1} &=&8\left( R^{4}+1\right) \left(
R^{8}-12R^{7}+R^{6}+8R^{5}+R^{4}-20R^{3}-R^{2}+6\right) , \\
d_{1} &=&-16\left( R-1\right) \left( R^{3}+R^{2}+R-1\right) \left(
R^{4}+1\right) ^{2},
\end{eqnarray*}%
$A_{0}=a_{0}+b_{0}h+c_{0}h^{2}+d_{0}h^{3}$:%
\begin{eqnarray*}
a_{0} &=&\left( R-1\right) \left( R^{4}+4R^{2}+4R+3\right) \left( 
\begin{array}{c}
R^{7}+7R^{6}+9R^{5}-13R^{4} \\ 
+7R^{3}+9R^{2}-R-3%
\end{array}%
\right) , \\
b_{0} &=&-4\left( R-1\right) \left( 
\begin{array}{c}
3R^{10}+9R^{9}-12R^{8}+25R^{7}+29R^{6}+15R^{5} \\ 
-9R^{4}+23R^{3}+24R^{2}+8R-3%
\end{array}%
\right) , \\
c_{0} &=&-4\left( R-1\right) \left( R^{4}+1\right) \left( 
\begin{array}{c}
R^{7}+3R^{6}-15R^{5}+11R^{4} \\ 
+3R^{3}-7R^{2}-13R+1%
\end{array}%
\right) , \\
d_{0} &=&16R\left( R-1\right) ^{2}\left( R^{4}+1\right) ^{2}.
\end{eqnarray*}

Let us stress here that acceptable solution for the time of the first impact
must fulfill consistency condition $T_{1}\equiv X\in \left( 0,h\right) $.
Now it follows that necessary conditions for existence of this solution can
be formulated. Indeed, the condition $A_{0}=f\left( R,h\right) =0$
guarantees existence of solution $T_{1}\equiv X=0$. Furthermore, after
change of variable $X=\tilde{X}+h$ the equation (\ref{poly}) is written as $%
\tilde{A}_{6}\tilde{X}^{6}+\ldots +\tilde{A}_{1}\tilde{X}+\tilde{A}_{0}=0$
and the condition $\tilde{A}%
_{0}=A_{6}h^{6}+A_{5}h^{5}+A_{4}h^{4}+A_{3}h^{3}+A_{2}h^{2}+A_{1}h+A_{0}=0$
guarantees existence of the solution $\tilde{X}=0$ and hence existence of
solution $T_{1}\equiv X=h$. Region of acceptable values of parameters $R$, $%
h $ is shown in Fig. 6 - it is placed between thin solid lines (which
correspond to the condition $T_{1}=h$) and below medium solid line (the
condition $T_{1}=0$).

\begin{figure}[h]
\begin{equation*}
\includegraphics[width=7 cm]{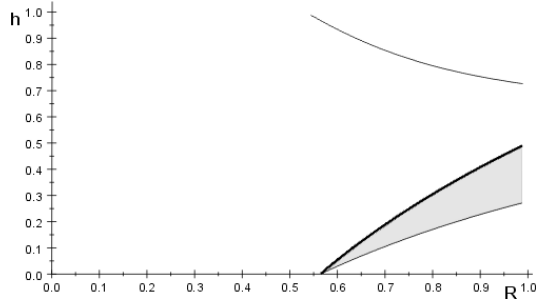}
\end{equation*}%
\caption{Acceptable parameter values (shaded region).}
\label{F6}
\end{figure}

The solution of Eq. (\ref{hom1}) is unstable and leads to a homoclinic--type
orbit and thus can be referred to as the homoclinic cycle (see \cite%
{Devaney2003} for the definition of a homoclinic point and a homoclinic
orbit). Indeed, for the initial condition $1-\epsilon _{1}<T_{2}<1$ the
orbit is attracted by the $2^{2}$ - cycle $T_{\ast 1},T_{\ast 2}=1,T_{\ast
3},T_{\ast 4}$, while for $0<T_{2}<\epsilon _{2}$ the fixed point $T_{\ast
2}=1$ is repelling, but the orbit returns eventually to the $2^{2}$ - cycle
(provided that a coexisting attractor does not capture the trajectory). We
shall compute the repelling branch in the next Section.

\section{Computing the homoclinic orbit}

To analyse structure of the chaotic attractor shown in Fig. 2 we have solved
Eq. (\ref{hom1}) for $R=0.85$, $h=0.2623$, computing thus critical value of
the parameter $\gamma $ and the homoclinic cycle where $\left( T_{\ast
2},V_{\ast 2}\right) $ is the homoclinic point:%
\begin{equation}
\begin{array}{ll}
\gamma _{cr}^{\left( 1\right) }=0.090\,581\,947\,119\,, &  \\ 
T_{\ast 1}=0.122\,922\,181\,823\,, & V_{\ast 1}=0.828\,678\,917\,871\,, \\ 
T_{\ast 2}=1^{-}\,, & V_{\ast 2}=0.559\,494\,302\,251\,, \\ 
T_{\ast 3}=0.\,214\,157\,103\,932\,, & V_{\ast 3}=0.527\,370\,736\,661\,, \\ 
T_{\ast 4}=0.827\,577\,663\,824\,, & V_{\ast 4}=0.367\,388\,917\,196\,.%
\end{array}
\label{homP}
\end{equation}

This solution is attracting for initial condition $T_{2}\in \left(
1-\epsilon _{1},1\right) $ and $V_{2}\cong V_{\ast 2}$. The sequence $%
\left\{ T_{i},V_{i}\right\} $ starting from such initial condition belongs
to the attracting branch of the homoclinic orbit. The repelling branch is
obtained in the following way. We start from $T_{1}=T_{\ast 1}$, $%
V_{1}=V_{\ast 1}$. $T_{2}$ and $V_{2}$ are computed from the following
equations:%
\begin{eqnarray}
\frac{\gamma }{h}\left( T_{2}-1\right) &=&\frac{\gamma }{h}T_{\ast 1}-\left(
T_{2}-T_{\ast 1}\right) ^{2}+\left( T_{2}-T_{\ast 1}\right) V_{\ast 1}
\label{homRa} \\
V_{2} &=&-RV_{\ast 1}+2R\left( T_{2}-T_{\ast 1}\right) +\left( 1+R\right) 
\frac{\gamma }{h}  \label{homRb}
\end{eqnarray}

The solution of the first equation is of course $T_{2}=1$. We assume now in (%
\ref{homRb}) that at the impact the table is just about going up with
velocity $\dot{Y}\left( T_{2}\right) =1/h$ rather than it has just finished
going down with velocity $\dot{Y}\left( T_{2}\right) =-1/\left( 1-h\right) $
(therefore this equation differs from the second of equations in (\ref{hom1}%
)). We thus compute from Eq. (\ref{homRb}), for $R=0.85$, $h=0.2623$ and $%
\gamma =\gamma _{cr}^{\left( 1\right) }=0.090\,581\,947\,119$, that $%
V_{2}=1.\,425\,529\,027\,600$ (let us stress again that using the second of
Eqs. (\ref{hom1}) we get $V_{\ast 2}=0.559\,494\,302\,251$)$\,$. Due to
symmetry of the dynamics $T_{i}\rightarrow T_{i}+1$ the first point of the
repelling branch of the homoclinic orbit can be assumed as $T_{1}^{R}=0$, $%
V_{1}^{R}=1.\,425\,529\,027\,600$ .

We have thus computed numerically the repelling branch of the homoclinic
orbit starting from the initial condition $T_{1}^{R}$, $V_{1}^{R}$ ($R=0.85$%
, $h=0.2623$, $\gamma =\gamma _{cr}^{\left( 1\right) }=0.090\,581\,947\,119$%
).

\begin{figure}[h]
\begin{equation*}
\includegraphics[width=7 cm]{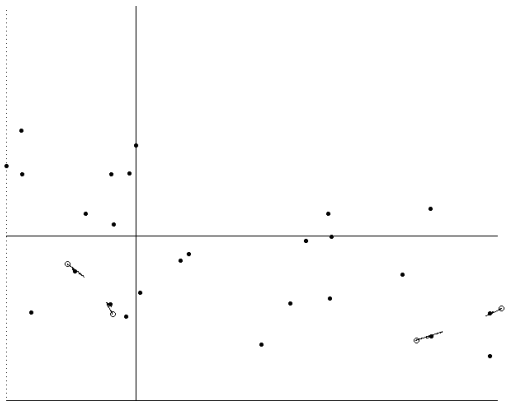}
\end{equation*}%
\caption{The homoclinic orbit.}
\label{F7}
\end{figure}

We have shown in Fig. 6 first twenty six points (full circles) of the
repelling branch of the homoclinic orbit starting from the point $\left(
T_{1}^{R},V_{1}^{R}\right) $ - the outermost full circle in the figure,
lying on the vertical axis. These points agree very well with positions of
twenty six clouds of points belonging to the chaotic attractor shown in Fig.
2, computed for $\gamma =0.09059$. The next points (dots) of the repelling
branch of the homoclinic trajectory enter four connected parts placed as in
Fig. 2 and tend, as an attracting branch, to the homoclinic cycle (larger
open circles) containing the homoclinic point $\left( T_{\ast 2},V_{\ast
2}\right) $ - the outermost open circle.

This homoclinic structure is preserved in the interval $\gamma _{cr}^{\left(
1\right) }=0.0905819471\ldots <\gamma <0.09060$, $R=0.85$, $h=0.2623$\ until
this attractor is substituted by a new one due to crisis (the unstable cycle
collides with one of clouds of points belonging to the attractor), see the
bifurcation diagram, Fig. 1, and Figs. 2,3.

\section{Summary and discussion}

We have found a generic scenario of transition to chaos for dynamics of a
ball moving vertically in gravitational field and colliding with a table
moving vertically with piecewise constant velocity.

According to this scenario a periodic and stable solution is destroyed via a
corner bifurcation \cite{diBernardo2008} in a corner event, $T_{\ast
i}=T_{cr}^{\left( 1\right) }=h$ or $T_{\ast i}=T_{cr}^{\left( 2\right) }=1$.
In the present paper the solution, defined analytically by $T_{\ast 2}=1$ in
Eq. (\ref{equations}), is a homoclinic--type orbit and leads to mixing and
hence to chaotic dynamics. This homoclinic--type orbit is untypical in the
sense that it is not a saddle structure but its origin is related to
discontinuous dynamics in the neighborhood of $T_{cr}^{\left( 2\right) }$.

\end{document}